\documentclass[12pt]{article}

\usepackage{amsmath}
\usepackage{amssymb}
\usepackage{amsfonts}
\usepackage[colorlinks=true,linkcolor=RawSienna,citecolor=RawSienna,urlcolor=RawSienna,bookmarksopen=true,pdfstartview=FitB]{hyperref}
\usepackage[dvipsnames]{xcolor}
\usepackage[top=1.1in, bottom=1.1in, left=1in, right=1in]{geometry}
\usepackage[onehalfspacing]{setspace}
\usepackage{natbib}

\usepackage{algorithm}
\usepackage{algpseudocode}
\usepackage{graphicx}
\usepackage{caption}
\usepackage{subcaption}
\usepackage{enumerate}

\usepackage{bm}

\newcommand{\prob}{\mathbb{P}}

\begin{document}

\title{\vspace{-1.2cm}  Discovering Candidate Genes Regulated by GWAS Signals in Cis and Trans}
\author{Samhita Pal$^1$ and Xinge Jessie Jeng$^{1^*}$ \\
{\large $^1$Department of Statistics, North Carolina State University}}
 \date{*Corresponding author. Email: xjjeng@ncsu.edu \\
	Contributing author: spal4@ncsu.edu}
\maketitle

\begin{abstract}
 Understanding the genetic underpinnings of complex traits and diseases has been greatly advanced by genome-wide association studies (GWAS). However, a significant portion of trait heritability remains unexplained, known as ``missing heritability". Most GWAS loci reside in non-coding regions, posing challenges in understanding their functional impact. Integrating GWAS with functional genomic data, such as expression quantitative trait loci (eQTLs), can bridge this gap. This study introduces a novel approach to discover candidate genes regulated by GWAS signals in both cis and trans. Unlike existing eQTL studies that focus solely on cis-eQTLs or consider cis- and trans-QTLs separately, we utilize adaptive statistical metrics that can reflect both the strong, sparse effects of cis-eQTLs and the weak, dense effects of trans-eQTLs. Consequently, candidate genes regulated by the joint effects can be prioritized. We demonstrate the efficiency of our method through theoretical and numerical analyses and apply it to adipose eQTL data from the METabolic Syndrome in Men (METSIM) study, uncovering genes playing important roles in the regulatory networks influencing cardiometabolic traits. Our findings offer new insights into the genetic regulation of complex traits and present a practical framework for identifying key regulatory genes based on joint eQTL effects.

\medskip


\end{abstract}


\section{Introduction}

A central goal of genetics is to understand how genetic variation,  environment factors, and other sources of variations influence traits. 
Over the past decades, genome-wide association studies (GWASs), facilitated by high-throughput sequencing technologies, have revolutionized human genetics.  
However, GWAS have also left several outstanding questions: the identified GWAS loci have explained only a small percentage of trait heritability, known as the ``missing heritability" problem; as sample sizes for GWAS continue to grow, the effect sizes of newly identified loci are mostly very weak; and the majority of GWAS loci are located in non-coding putatively regulatory regions of the genome, leaving it unclear which genes they regulate and how these genes function in pathways to impact the traits. 
These difficulties in interpreting GWAS results 
have hindered our efforts to understand disease etiologies and develop new therapies \citep{visscher2017, cano2020gwas}. 

To tackle these challenges, emerging studies  integrate GWAS results with functional genomic data such as gene expression or chromatin activity profiles assayed across a range of cell types and tissues.
Many large consortia, such as GTEx \citep{gtex2015genotype}, have generated gene expression data for different tissues and cell lines together with genotype information. Such data have been extensively analyzed to study the regulatory effects of SNPs, i.e., expression quantitative trait loci (eQTLs), on gene expression across human tissues, and the results have proven informative for GWAS analysis.   
For example, transcriptome-wide association studies (TWAS) and colocalization analysis have been developed in recent years to identify genes causally involved in complex diseases based on GWAS and eQTL results. 
TWAS leverages information from eQTL catalogs to impute gene expression values and directly infer the association between traits and the imputed gene expression for the gene under investigation \citep{gusev2016integrative, gamazon2015gene}. In contrast, colocalization analysis integrates association results from GWAS and eQTL mapping to identify instances where both traits share a causal variant. The rationale is that if a GWAS trait and a gene's expression share the same associated SNP, it may suggest a regulatory (and thus putative causal) role of the SNP mediated through the gene on the GWAS trait \citep{cannon2018deciphering, deng2020powerful}.

Integrative analyses provide important observations that for most complex traits, the most significant signals are located near genes that make functional sense, although those signals and genes contribute only a small fraction of the total heritability. In eQTL analysis, such nearby signals are called cis-acting eQTLs (cis-eQTL).  Because each cis-eQTL influences only one phenotype (the local gene), they cannot elucidate the co-regulatory landscape of the transcriptome and thus the higher organization of gene regulation \citep{westra2013systematic, bryois2014cis, smemo2014obesity, claussnitzer2015fto, brynedal2017large}.

It has been observed that heritability is dominated by variation in a large number of non-coding common variants spread broadly across the genome and mediated through a wide range of gene functional categories. Such distant variants are called trans-acting eQTLs (trans-eQTLs) \citep{price2008effects, zhu2016integration, yao2017dynamic, civelek2017genetic}. 
An emerging theory known as the omnigenic model suggests that complex traits or diseases are governed by possibly all genes expressed in relevant cells through network behaviors. Among these genes, there is a set of ``core genes" directly involved in the biological pathways for each trait and a larger number of ``peripheral genes" that can only affect the trait indirectly by connecting with these core genes via regulatory networks.  The omnigenic model indicates that cis and trans effects of trait-associated variation are mediated by core genes, which are directly involved in the biological pathways for the trait. Moreover, it is discussed that for typical genes, the majority of expression heritability is caused by trans-eQTLs \citep{boyle2017expanded, liu2019trans}.  

However, for purely practical reasons, most eQTL mapping studies to date have concentrated on cis-eQTLs. This is because trans-eQTLs often have small regulatory effects in humans, making them notoriously difficult to identify. Analyses involving trans-eQTLs often require substantially large sample sizes, significant computational resources, and extremely stringent multiplicity adjustments.

In this paper, we propose a method to directly identify expression genes (eGenes) whose heritability is influenced by GWAS signals in both cis and trans, bypassing the need to identify weak trans-eQTLs individually. Our approach uses statistical metrics sensitive to both cis- and trans-acting signals -- cis signals being relatively strong but sparse, and trans signals being relatively weak but dense. The eGenes prioritized by these metrics can offer valuable insights into the mechanisms underlying disease-related traits and may serve as candidates for core genes in omnigenic models.

The statistical metrics utilized in our method for eGene discovery have been studied in high-dimensional inference and discovered as being optimally adaptive for heterogeneous and heteroscedastic mixture model detection \citep{donoho2004higher, jager2007goodness, cai2011optimal, jeng2013simultaneous, jeng2020effective}. 
In the context of eQTL analysis, we investigate their adaptivity to both cis- and trans-acting signals, providing theoretical insights into their efficiency. Numerical analyses are conducted to highlight the advantages of our method in discovering three types of eGenes: those regulated exclusively by cis-eQTLs, those regulated solely by trans-eQTLs, and those influenced by both.  We apply this new method to analyze adipose eQTL data from the METabolic Syndrome in Men (METSIM) study. The results provide valuable insights into the genes that play important roles in the regulatory networks for cardiometabolic traits.

\section{Methodology}


\subsection{Formulation}

Suppose there are $m$ SNPs, which are identified through GWAS to be associated with at least one of the $q$ disease-related traits under consideration. We consider all $K$ genes as potentially regulated by one or more of the $m$ GWAS signals. We obtain the summary statistics $Z_{jk}$, $1 \le j \le m$, $1 \le k \le K$, from marginal association tests between the $m$ SNPs and the $K$ genes. For a given gene $k$, assume that 
\[
Z_{jk} \sim F_{k,0} \cdot 1\{j \in I_{k,0}\} + F_{k,1} \cdot 1\{j \in I_{k,1}\}, \qquad 1 \le j \le m, 
\]
where $I_{k,0}$ and $I_{k,1}$ represent the set of indices for noise variants and eQTL signals, respectively, which vary from gene to gene. Moreover, $F_{k,0}$ and $F_{k,1}$ represent the noise and signal distribution of $Z_{jk}$, respectively, which also vary from gene to gene. In this model, only the null distribution $F_{k,0}$ is specified. The other components, $I_{k,0}$, $I_{k,1}$, and $F_{k,1}$, are unknown. 

Specifically, we expect that for a gene $k$ regulated solely by  cis-eQTLs, its $I_{k,1}$ set should be small, and its signal distribution $F_{k,1}$ would concentrate more on strong intensities. On the other hand, for a gene $k'$ regulated by  multiple trans-eQTLs, its $I_{k',1}$ set should be relatively larger, and its signal distribution would concentrate more on weak intensities. For a gene influenced by both cis- and trans-eQTLs, its signal distribution could span both strong and weak intensities. Our goal is to discover all these different types of eGenes by prioritizing them over irrelevant genes and identify them with statistical guarantees.  

\subsection{Adaptive Metrics and Algorithm} \label{sec:metric_algorithm}

A key component of our proposed eGene discovery method is the use of a statistical metric that has been identified as optimally adaptive for detecting various types of mixture models. This metric belongs to a general family of goodness-of-fit (GOF) tests, known for demonstrating asymptotic optimality in detecting weak and rare signals amidst a large number of noise variables  \citep{donoho2004higher, jager2007goodness, donoho2008higher, cai2011optimal, jeng2013simultaneous, jeng2020effective}. Two members in the GOF family, the Higher Criticism (HC) test and the Berk-Jones (BJ) test, have also been extensively studied in the literature. Specifically, the HC test is renowned for its effectiveness in identifying rare signals, while the BJ test is recognized for its robustness across different signal patterns in finite samples \citep{li2015higher, barnett2017generalized, zhang2020distributions, sun2020genetic}. 

The HC and BJ statistics are defined as follows: For a gene $k$, derive the $p$-values $\{p_{jk}\}_{j=1, \ldots, m}$ of $\{Z_{jk}\}_{j=1, \ldots, m}$;  order the $p$-values increasingly so that $p_{(1),k} \le p_{(m),k} \le \ldots, \le p_{(m),k}$;  and define
\begin{equation} \label{def:HC_stat}
HC_k = \sup_{1\le j \le m} \sqrt{m} \frac{j/m - p_{(j),m}} {\sqrt{p_{(j),m}(1 - p_{(j),m})}}
\end{equation}
and 
\begin{equation} \label{def:BJ_stat}
BJ_k = \sup_{1 \le j \le m/2}\left\{ {j\over m} \log ({j/m \over p_{(j),k}})  + (1-{j\over m}) \log ({1-j/m \over 1-p_{(j),k}})\right\}.
\end{equation}
Both HC and BJ statistics were designed to detect departures from a specified null hypothesis that all the data points $\{p_{jk}\}_{j=1, \ldots, m}$ follow the null distribution U$(0,1)$ \citep{berk1979goodness, donoho2004higher}. These statistics compare the ordered $p$-values $p_{(j),k}$ to their expected values $j/m$ under the null hypothesis and assess the maximum deviation of a weighted function of the differences between the observed and expected $p$-values. 
The weights are chosen to give more importance to the regions where the differences are most significant. Larger values of the statistics indicate stronger evidence that the data come from a mixture of noise and signal distributions rather than just the noise distribution.

Next, we outline the detailed steps for the proposed eGene discovery method, which utilizes either HC or BJ statistic as the adaptive metric ($T_k$) to capture the joint regulatory effects of the GWAS signals in cis and trans. The genes are ranked based on this adaptive metric, and candidate eGenes are identified with statistical significance. 

\noindent \underline{Step-by-Step Procedure}: 

\begin{enumerate}
	\item Identify the $m$ GWAS signals for a set of $q$ traits related to a specific disease.
	\item Obtain the $p$-values for the marginal association tests between the $m$ GWAS signals and all the $K$ genes under consideration. 
	\item Calculate the adaptive metrics $T_k, k=1,\ldots, K$, for all the genes using either (\ref{def:HC_stat}) or (\ref{def:BJ_stat}).  
	\item Rank the genes according to the values of $T_k$'s from largest to smallest.
	\item Calculate the $p$-values of $T_k$ for all $k =1, \ldots, K$. 
	\item Select candidate eGenes with significant $T_k$ $p$-values through multiple testing.  
\end{enumerate}

\noindent Remark: The above procedure is completely data-driven. We utilize the R package {\it SetTest} from \cite{zhang2020distributions} to calculate $HC_k$ and $BJ_k$ as defined in (\ref{def:HC_stat}) - (\ref{def:BJ_stat}) and to derive their $p$-values. This computation is fast and scalable for large-scale eGene discovery.  

The effectiveness of the proposed method hinges on the adaptivity of the $T_k$ statistic to both cis and trans-acting signals, for which we provide some theoretical insight in Section \ref{sec:theory} and numerical justifications in Section \ref{sec:simulation}.

\subsection{Theoretical Insight} \label{sec:theory}

For a given gene $k$, the problem of detecting the existence of any eQTL signals can be formulated  as a global hypothesis testing problem. 
Let $F_{k,0} = N(0, 1)$, representing the noise distribution after data standardization, and $F_{k,1} = N(\mu_k, \tau_k^2)$, representing the signal distribution, where both $\mu_k$ and $\tau_k^2 $ are unknown. Then, an irrelevant gene that is not influenced by any of the SNPs corresponds to the global null hypothesis   
\begin{equation} \label{def:H0}
H_{k,0}: Z_{jk} \sim N(0, 1),  \qquad 1 \le j \le m;
\end{equation}
and a candidate gene regulated by one or more of the SNPs corresponds to the alternative hypothesis
\begin{equation} \label{def:H1}
H_{k,1}: Z_{jk} \sim (1-\pi_k)N(0, 1) + \pi_k N(\mu_k, \tau_k^2), \qquad 1 \le j \le m,
\end{equation}
where $\pi_k$ represents the proportion of eQTL signals among all the SNPs, which varies from gene to gene. 

To gain theoretical insight into the effect of eQTL signal proportion, as well as the impact of signal mean and variance on the global testing problem, we re-parameterize some model parameters as in \cite{cai2011optimal}. This re-parameterization allows their effects to be represented on comparable scales and analyzed jointly. Specifically, let 
\begin{equation} \label{def:gamma}
	\pi_k = m^{-\gamma_k}, \qquad \gamma_k \in (0, 1),
\end{equation}
so that the signal sparsity is represented through the parameter $\gamma_k$ in a constant scale with $\gamma_k=0$ representing the extremely dense case where all SNPs are eQTL signals and  $\gamma_k=1$ representing the extremely sparse case where there is only one eQTL signal. Then, depending on the value of $\gamma_k$, we consider two scenarios: the sparse scenario with  $\gamma_k \in (1/2, 1]$ and the dense scenario with $\gamma_k \in (0, 1/2]$. In these two scenarios, the signal mean value is re-parameterized differently:
\begin{equation} \label{def:h}
	\mu_k = \begin{cases}
		\sqrt{2h_k \log m}, &   \gamma_k \in (1/2, 1), \\
		m^{-h_k},  &  \gamma_k \in (0, 1/2], 
	\end{cases}
\end{equation}
where $h_k>0$ is at the constant scale and represents the signal mean effect. The above re-parameterization reflects two different types of signals, one being sparse ($\gamma_k \in (1/2, 1)$) and relatively strong ($\mu_k=\sqrt{2h_k \log m}$), the other being dense ($\gamma_k \in (0, 1/2)$) and relatively weak ($\mu_k=m^{-h_k}$).

After the re-parameterization, all the parameters $\gamma_k$, $h_k$, and $\tau_k$ are at a constant scale, which facilitates the analysis of their joint effects. 
Notably, the detection boundary $\rho(\gamma_k,\tau_k)$ for the global testing problem has been found in \cite{cai2011optimal} and \cite{jeng2013simultaneous}, assuming independence among $Z_{jk}, 1\le j\le m$. Its specific form varies across different scenarios as follows:
 
When signals are relatively sparse with $\gamma_k \in (1/2, 1)$, 
\begin{equation} \label{def:rho_sparse}
	\rho(\gamma_k,\tau_k) = \begin{cases}
		(2 - \tau^2_k)\left(\gamma_k - 1/2 \right), &  \tau^2_k \in [1,~2) \text{~~and~~} \gamma_k \in (1/2,~ 1-\tau^2_k/4],  \\
		\left( 1 - \tau_k \sqrt{1 - \gamma_k} \right)^2, & \tau^2_k \in [1,~2) \text{~~and~~} \gamma_k \in  (1 - \tau^2_k/4, ~ 1),  \\
		0, & \tau^2_k>2 \text{~~and~~} \gamma_k \in (1/2,~ 1 - 1/\tau^2_k],  \\
		\left( 1 - \tau_k \sqrt{1 - \gamma_k} \right)^2, & \tau^2_k>2 \text{~~and~~} \gamma_k \in (1 - 1/\tau^2_k,~ 1).
	\end{cases}
\end{equation}

When signals are relatively dense with  $\gamma_k \in (0, 1/2]$, 
\begin{equation} \label{def:rho_dense}
	\rho(\gamma_k,\tau_k) = \begin{cases}
		\infty, & \tau^2_k >1 \text{~~and~~} \gamma_k \in (0, 1/2], \\
		1/2 - \gamma_k, & \tau^2_k = 1 \text{~~and~~} \gamma_k \in (0, 1/2].
	\end{cases}
\end{equation}    

The detection boundary results can serve as a benchmark for method evaluation as it implies the following three aspects: 

\begin{enumerate}
	\item  When the signal mean value does not reach the detection boundary, i.e.,  
	\begin{equation} \label{def:undet_region}
		h \begin{cases}
			< \rho(\gamma_k, \tau_k), & \gamma_k \in (1/2, 1), \\
			> \rho(\gamma_k, \tau_k), & \gamma_k \in (0, 1/2],
		\end{cases}
	\end{equation}
	no methods can consistently separate $H_{k,1}$ from $H_{k,0}$, i.e., 
	\[
	\prob_{H_{k,0}} \left( \text{Reject } H_{k,0}\right) + \prob_{H_{k,1}} \left( \text{Fail to reject } H_{k,0}\right) \rightarrow 1 \quad \text{ as } \quad m \rightarrow \infty. 
	\]
	\item If a method can consistently separate $H_{k,1}$ from $H_{k,0}$ with 
	\[
	\prob_{H_{k,0}} \left( \text{Reject } H_{k,0}\right) + \prob_{H_1} \left( \text{Fail to reject } H_{k,0}\right) \rightarrow 0 \quad \text{ as } \quad m \rightarrow \infty
	\]
	whenever the signal mean value passes the detection boundary, i.e., 
	\begin{equation} \label{def:det_region}
		h \begin{cases}
			> \rho(\gamma_k, \tau_k), & \gamma_k \in (1/2, 1], \\
			< \rho(\gamma_k, \tau_k), & \gamma_k \in (0, 1/2],
		\end{cases}
	\end{equation}
	then this method is considered an optimal method. 
	\item If a method can achieve optimality and its implementation does not require specified values of the model parameters, then the method is considered an optimally adaptive method.
\end{enumerate}

In the literature, it has been proved that all the members in the aforementioned GOF family are optimally adaptive for the global testing problem in the sparse scenario ($\gamma_k \in (1/2, 1)$) with equal variance ($\tau_k^2 = 1$)\citep{jager2007goodness}. Moreover, the optimal adaptivity of HC test has also been proved in the extended scenarios with  ($\gamma_k \in (0, 1]$) and possibly unequal variance ($\tau_k^2 \ge 1$) \citep{cai2011optimal, jeng2013simultaneous}. 
We expect that the BJ test performs comparatively to the HC test in the extended scenarios, and this conjecture is supported in the following numerical studies. 

\section{Simulation} \label{sec:simulation}

\subsection{Simulation Setup}

\begin{figure}[h]
	\centering
	\includegraphics[width = 0.8 \textwidth]{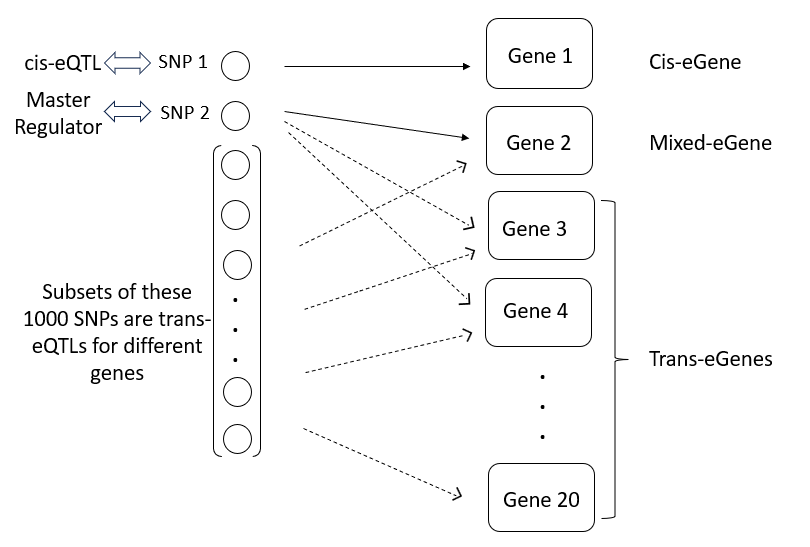}
	\caption{Three types of eGenes and their regulatory eQTLs:  The solid lines represent cis effect and the dotted lines represent trans effect of the corresponding SNPs. Gene 1 is a cis-eGene regulated by SNP 1 in cis; Gene 2 is a mixed-eGene regulated by SNP 2 in cis and a subset of SNPs in trans; Genes 3 - 20 are trans-eGenes regulated by various subsets of SNPs in trans.}
	\label{fig:eQTL_map}
\end{figure}

The simulation study aims at checking the competence and efficiency of the proposed eGene discovery method. 
In the study, we consider $100$ genes and $8657$ SNPs. 
We design three types of eGenes as depicted in Figure \ref{fig:eQTL_map}:  \\
\underline{\it Cis-eGene}: illustrated by Gene 1,  which is only regulated by one cis-eQTL (SNP 1). \\
\underline{\it Mixed-eGene}: illustrated by Gene 2, which is co-regulated by one cis-eQTL (SNP 2) and a set of trans-eQTLs (a random subset from SNPs 3 - 1002). \\
\underline{\it Trans-eGene}: illustrated by Genes 3 - 20, each of which are regulated by a set of trans-eQTLs  (random subsets from SNPs 3 - 1002). \\
The remaining genes (Genes 21 - 100) are not associated with any of the SNPs and are not depicted in Figure \ref{fig:eQTL_map}. 

We generate the summary statistics  $Z_{jk}$ for the association tests between the 100 genes and $8657$ SNPs as follows: For $k = 1, \ldots, 100$, let 
\[
Z_{jk} \sim N(0, 1) \cdot 1\{j \in I_{k,0}\} + N(\mu_k, \tau^2_k) \cdot 1\{j \in I_{k,1}\}, \qquad j =1, \ldots, 8657.
\]
For $k=1$ (Gene 1), $I_{1,1} = \{1\}$ and $\mu_1=A$, representing the effect of one strong cis-eQTL.   
For $k=2$ (Gene 2), $I_{2,1} = \{2\} \cup S_2$ and $\mu_2 = A \cdot 1\{j=2\} + B \cdot 1\{j \in S_2\}$, representing the mixed effect of one strong cis-eQTL and many weak trans-eQTLs. For $k= 3 - 20$ (Genes 3 - 20), $I_{k,1} = S_k$ and $\mu_k = B \cdot 1\{j \in S_k\}$, representing the effect of many weak trans-eQTLs. Note that $S_2, S_3, \ldots, S_{20}$ are randomly selected subsets of SNPs from SNPs 3 - 1002 with a common cardinality $s = |S_k|$. For $k=21 - 100$ (Genes 21 - 100), $I_{k, 1} = \emptyset$, so that they are irrelevant genes. 
More specific values of the model parameters in the simulation examples include $A=4$, $B=0$ or 0.2, $\tau^2_k =1.05$ or 1.5, and  $s=500$ or 50. Note that $B=0$ represents the case where the weak trans-eQTLs barely show a non-zero mean effect, and their effects are solely demonstrated through increased variability ($\tau_k^2>1$).

\subsection{Methods Comparison}

We apply the proposed eGene discovery procedure, as described in Section \ref{sec:metric_algorithm}, to the simulated data. Here, we focus on evaluating the efficiency of the proposed method in prioritizing true eGenes over irrelevant ones. We compare both the HC version and the BJ version of our procedure with other existing methods that use different information pooling strategies to summarize the eQTL effects received by a gene. These methods are summarized as follows: 
\begin{enumerate}[{(a)}]
	\item \underline{\it Mean-based method}: For a given gene k, summarize the eQTL effects it receives from all the SNPs through the mean value  $\sum_{j=1}^m Z_{jk}/ m$. 
	\item \underline{\it MinimumP-based method}: For a given gene k, summarize the eQTL effects through the minimum $p$-value of $Z_{jk}$: $\min_{1 \le j \le m} p_{jk}$.
	\item \underline{\it HC-based method}: Use the HC statistic in (\ref{def:HC_stat}) to summarize the eQTL effects for each gene. 
	\item \underline{\it BJ-based method}: Use the BJ Statistic in (\ref{def:BJ_stat}) to summarize the eQTL effects for each gene. 
\end{enumerate}

Both the Mean- and MinimumP-based methods are commonly used approaches for information pooling. 
We compare the finite-sample performances of these methods in simulation settings designed to reflect different types of eGenes. 
The comparisons focus on the ability of these methods to prioritize different types of eGenes (Genes 1–20) over the irrelevant genes (Genes 21–100). 

The results are presented using the Precision-Recall (PR) curve.  The PR curve is particularly effective for evaluating the performance of a method in prioritizing relevant variables over irrelevant variables or, in other words, the ranking efficiency of a method. Here, the four methods in (a) - (d) rank the genes differently according to different information pooling approaches. To evaluate the ranking efficiency of these methods, a series of two metrics: Precision and Recall, are calculated along the ranked genes for each method. That is, for a list of ranked genes by a method, calculate 
\[
\text{Precision}_r = \frac{\text{true eGenes in the top r genes}}{\text{r}}, \quad 
\text{Recall}_r = \frac{\text{true eGenes in the top r genes}}{\text{all true eGenes}}, 
\]
for all $r =1, \ldots, K(=100)$. As the rank r increasing from 1 to 100, Recall$_r$ increases to reach 1 eventually while {Precision}$_r$ generally decreases. The PR curve is plotted with Recall$_r$ on the x-axis and Precision$_r$ on the y-axis, and the area under the PR curve (AUC-PR) is a single scalar value that summarizes the overall performance of the method in prioritizing true eGenes over irrelevant genes. Compared to the commonly used ROC curve, PR curve is more appropriate when dealing with imbalanced datasets where the number of signals is expected to be much less than that of the noise variables.

\subsection{Simulation Results}

\begin{figure}[htbp]\label{simu_2}
	\centering
	\begin{subfigure}[t]{0.49\textwidth}
		\includegraphics[width=\textwidth, height=6cm]{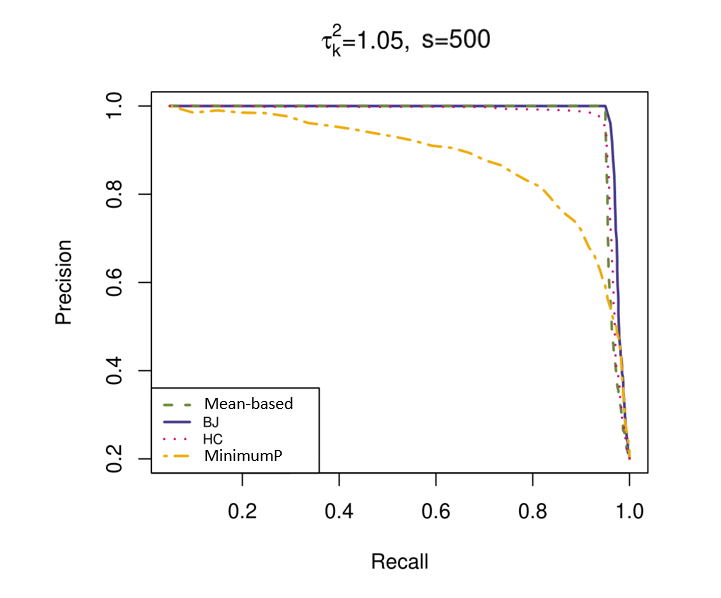}
	\end{subfigure}
	 \begin{subfigure}[t]{0.45\textwidth}
		\includegraphics[width=\textwidth, height=6cm]{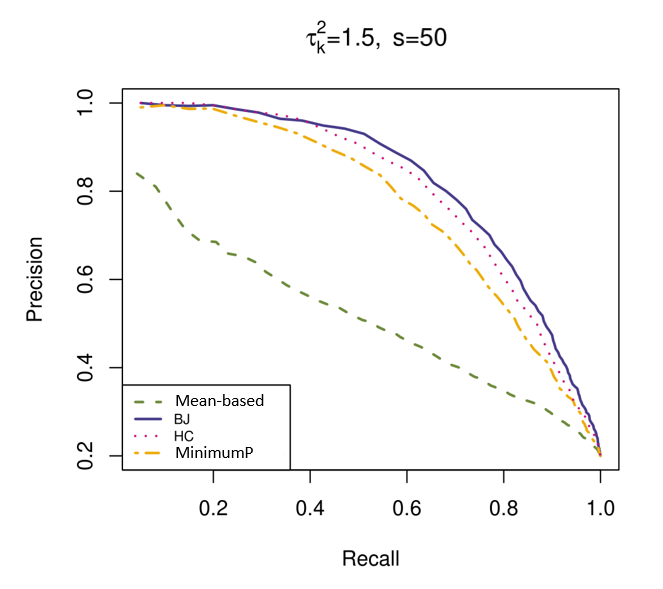}
	\end{subfigure}
\caption{PR Curves of all the methods. The Precision and Recall metrics are presented in mean values from 100 replications. There are 20 eGenes, including one cis-eGene, one mixed-eGene, and 18 trans-eGenes. The trans-eQTLs of an eGene include $s$ SNPs randomly selected from 1000 SNPs with possible overlapping across eGenes. The left panel has weak and dense trans-eQTL signals ($\tau_k^2=1.05, s=500$), whereas the right panel has stronger and sparser trans-eQTL signals ($\tau_k^2=1.5, s=50$)}.
\label{fig:sim_results}
\end{figure}

Figure \ref{fig:sim_results} shows that in the settings where different types of eGenes co-exist and the trans-eGenes are the majority, the MinimumP method performs the worst among the four methods in the left panel, where trans-eQTL signals are relatively weak and dense ($\tau_k^2=1.05, s=500$). Conversely, in the right panel where trans-eQTL signals are stronger and sparser ($\tau_k^2=1.5, s=50$), the Mean-based method performs the worst. In both scenarios, the BJ and HC methods exhibit more stable results and outperform the other two methods, with BJ and HC performing comparably.

Figure \ref{fig:sim_results} presents results with a zero mean value for trans-eQTLs ($B=0$). Similar results are observed with $B=0.2$ and are thus omitted to save space.

\section{Real Data Analysis} \label{sec:application}

We applied the proposed eGene discovery method to adipose eQTL data from the METSIM study \citep{laakso2017metabolic, raulerson2019adipose}.  This dataset consisted of RNA-seq data generated on subcutaneous adipose tissue biopsies from 426 males from Kuopio, Finland.  While \cite{raulerson2019adipose} primarily investigated adipose cis-eQTL ($\le$ 1Mb) colocalized with GWAS signals, we aimed here to discover eGenes regulated by GWAS variants in both cis and trans ($>$ 1Mb). The RNA-seq data quality control, normalization, and covariates were described in \cite{raulerson2019adipose}. Trans-eQTL analyses excluded pseudogenes and genes with low mappability, and we used 3 PEER factors as covariates for both cis and trans analyses. We obtained the $p$-values for the marginal association tests between the 2,879 GWAS signals (minor allele frequency $\ge 0.01$ and imputation $R2\ge 0.5$) and 24,371 genes described in that study. We found that a small number of genes with only one insignificant eQTL data point can cause errors when deriving their BJ p-values using the SetTest package. Consequently, we excluded these genes and conducted our analysis with the remaining 21,593 genes. 

As detailed in Section \ref{sec:metric_algorithm}, we calculated the BJ statistics and ranked the genes accordingly.
We further derived the BJ $p$-values of all the 21593 genes and applied multiple testing (Bonferroni $<0.01$) to select statistically significant eGenes. This resulted in the selection of 424 genes. The complete list of selected genes, ranked by their BJ statistics, is provided in the Appendix. 

Figure \ref{fig:data} presents the histogram of the BJ $p$-values of the 424 selected genes in the first panel. It can been seen that most of the selected genes have their BJ p-values much lower than the Bonferroni threshold (4.103237e-07). The results are supported by the outcomes of the HC method as shown in the second panel of Figure \ref{fig:data}, where the HC $p$-values of the selected 424 genes are also generally very small. 

\begin{figure}[htbp]
	\centering
	\includegraphics[width = 0.32 \textwidth]{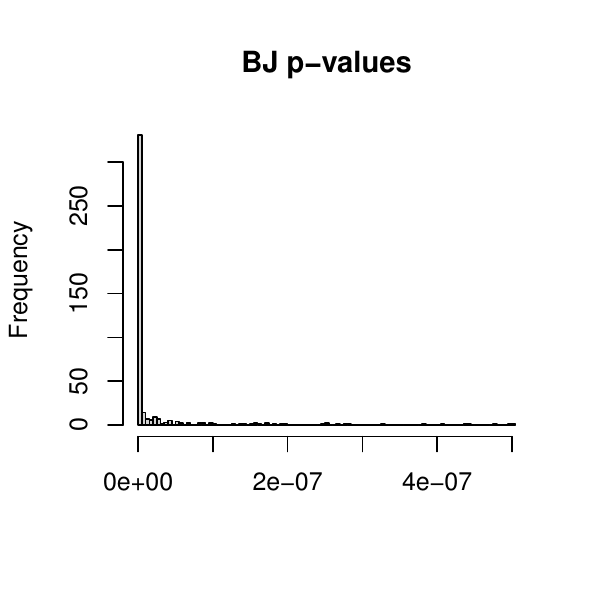}
	\includegraphics[width = 0.32 \textwidth]{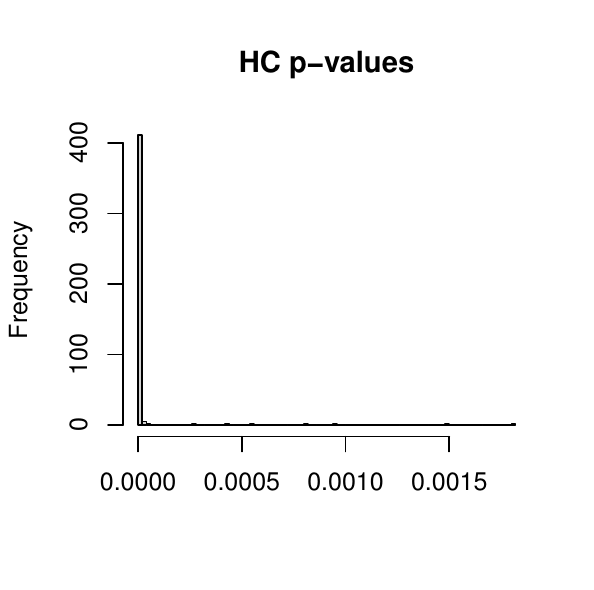}
	\includegraphics[width = 0.32 \textwidth]{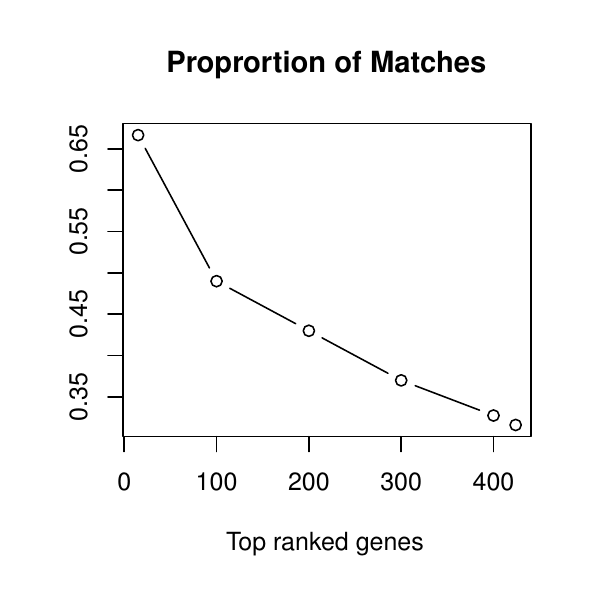}
	\caption{The first panel displays the histogram of BJ $p$-values for the 424 selected genes. The second panel presents the histogram of HC $p$-values of these genes. The third panel illustrates the proportion of matches between our ranked genes and those identified in \cite{raulerson2019adipose}.}
	\label{fig:data}
\end{figure}

We further compared our results with the eGenes identified in \cite{raulerson2019adipose}, where 318 eGenes (287 primary and 31 secondary) are identified using only the cis-eQTL data. We found 134 common genes between the two sets of results. As illustrated in the third panel of Figure \ref{fig:data}, the matching proportion along our list of ranked selected genes generally decreases from  0.67 to 0.32, indicating that our method effectively prioritize true eGenes over irrelevant ones. 

It is noteworthy that, different from  \cite{raulerson2019adipose}, our study simultaneously analyzes both cis and trans-eQTL data and captures the joint cis- and trans-acting effects on the genes.  
This approach could enhance the discovery of eGenes that play significant roles in encoding multifunctional proteins or regulating various cellular processes. The top 5 genes, as presented in Table \ref{tab:top_genes}, illustrate such examples.

\begin{table}[htbp]
	\centering
	\begin{tabular}{clll}
		\hline
		Rank & Gene Name &   Function Interpretation \\
		\hline
		1 & C1QTNF4
		& 
		Serum C1q/TNF-Related Protein 4 levels have been shown to be \\
		&&  lower in type 2 diabetes patients with nonalcoholic fatty liver \\
		&& disease (NAFLD) than those without NAFLD. \\
		\hline
		2 & WDR6 
		& 
		Upregulation of WD40 repeat-containing protein 6 levels \\
		&& has been reported to drive hepatic de novo lipogenesis \\
		&& during insulin resistance. \\ 
		\hline
		3 & AC034243.1
		& 
		This long noncoding RNA is transcribed in the antisense direction \\
		&&  from the transcription start site of CTNNA1, which encodes \\
		&&  catenin alpha 1 and has been implicated in cancer. \\
		\hline
		4 & HLA-DQA1
		&
		This component of the major histocompatibility complex plays a \\ 
		&& keyrole in the immune system. \\
		\hline
		5 & PEX6
		& 
		Peroxisomal biogenesis factor 6 plays a role in peroxisomal \\
		&& protein import. \\
		\hline
	\end{tabular}%
	\caption{Top 5 genes ranked by the significance of their BJ statistics. Gene names and functions were obtained by searching Gene resources at the National Center for Biotechnology Information; additional references include \cite{han2023serum}, \cite{yao2023upregulation}, and \cite{lobo2021cancer}.}
	\label{tab:top_genes}
\end{table}

\section{Conclusion and Discussion}

In this paper, we present a new approach to discovering candidate eGenes by detecting the joint effects of cis- and trans-eQTLs. Leveraging both cis and trans-eQTL data, our method offers a more comprehensive understanding of gene regulation mechanisms compared to traditional methods that consider these effects separately. The novel application of the statistically adaptive metrics enhances the prioritization of key regulatory genes involved in complex traits and diseases.

Applying our approach to adipose eQTL data from the METSIM study, we discovered eGenes regulated by GWAS signals in both cis and trans across a broad spectrum of 93 cardiometabolic traits. 
This analysis identified 424 significant eGenes, exceeding the number of genes identified by methods that focused solely on cis-eQTLs.
The top-ranked eGenes selected by our method play significant roles in encoding multifunctional proteins and regulating various cellular processes.
These findings underscore the potential of our eGene discovery method in uncovering complex genetic regulatory mechanisms. The identified eGenes provide valuable insights into the genetic architecture of complex traits and offer promising targets for further functional studies and therapeutic interventions.

\vspace{0.2in}

\noindent {\bf Acknowledgments.}  The authors would like to thank Dr. Sarah Brotman and Dr. Karen Mohlke for their assistance with accessing eQTL data from the METSIM study. The authors also express their gratitude to Dr. Zheyang Wu and Dr. Hong Zhang for their support with the {\it SetTest} software.

\noindent {\bf Declarations.} 

\noindent {\bf Data availability} The adipose eQTL data used in this paper is sourced from \cite{raulerson2019adipose} and is publicly accessible.

\section*{Appendix A: Complete list of selected genes} \label{sec:complete_list}

\begin{footnotesize}
\begin{verbatim}
 [1] C1QTNF4             WDR6                AC034243.1          HLA-DQA1            PEX6               
 [6] UHRF1BP1            ERAP2               HLA-C               HSD17B12            RBM6               
[11] CDK2AP1             HLA-DQB1            LINC00680           WNT6                CERS4              
[16] CDC37P1             AC016683.6          AP006621.1          PAX8                JMJD7              
[21] DALRD3              HLA-DQB1-AS1        XXbac-BPG254F23.6   MLXIPL              PM20D1             
[26] SNX10               CTD-2651B20.3       XXbac-BPGBPG55C20.1 RP11-282O18.3       NBPF3              
[31] NCKIPSD             RHD                 XKR9                ADORA2B             HLA-DRB1           
[36] NPC1L1              MIR1307             AP006621.5          XXbac-BPG299F13.17  ITGB6              
[41] AS3MT               GNL3                SPTBN5              MAN2C1              DISP2              
[46] C2orf74             C10orf32-ASMT       LITAF               RP11-217B7.2        RP11-378A13.1      
[51] NPC1                KIAA1841            KIAA1875            FN3KRP              GBAP1              
[56] RP11-65I12.1        JMJD7-PLA2G4B       XXbac-BPG248L24.12  RP11-73M18.9        ADH1A              
[61] ZSWIM7              C15orf38-AP3S2      ANAPC4              RP11-296I10.6       MED24              
[66] NPIPB7              RHCE                AC034220.3          RP11-109L13.1       ZNF600             
[71] CEP192              LILRA3              SERBP1P3            C12orf65            AC068039.4         
[76] RP11-1348G14.4      GNMT                RPL36P4             TIPARP              GTF2H1             
[81] SMIM19              AC145343.2          AOC1                EXOSC6              RP11-755F10.1      
[86] DGKQ                MAST4-AS1           RP11-624M8.1        STAG3L1             LILRB2             
[91] RFT1                AC018720.10         PLA2G4B             GSDMA               EIF3KP1            
[96] NR1H3               NT5DC2              NPIPA5              INO80E              PPID               
[101] RP11-57H14.2        SH2B1               MEGF9               ERAP1               NDUFAF1            
[106] CDK5RAP3            RP11-673E1.1        AP3S2               EIF3C               PIDD               
[111] AMT                 PLCE1               CDH13               PSMD5-AS1           MAP1LC3A           
[116] TRIP4               RP11-179B2.2        HLA-DRB6            SLC47A1             RP11-252K23.2      
[121] STRA13              CSNK2B              XXbac-BPGBPG55C20.2 ITIH4               RP11-529H20.6      
[126] SLC5A11             VEGFB               KIAA1683            ATXN2L              RPS26              
[131] C10orf32            AL049840.1          CTD-2303H24.2       RP4-712E4.1         RP11-244F12.3      
[136] USP32P2             MYEOV               AP006621.6          CTC-228N24.3        CLEC18A            
[141] RP11-69E11.4        TMEM180             SLC28A2             CRYZ                CEACAMP3           
[146] CCDC12              CHST8               NPIPA3              NMRAL1              RP11-148O21.6      
[151] APOB                NPIPB6              RP11-211G23.2       RP1-140A9.1         EMC1               
[156] MST1L               CCDC116             SRR                 PEPD                C15orf57           
[161] BMP8A               ARHGAP1             MAP3K11             SH3PXD2B            G6PC2              
[166] SCAPER              SPPL3               RP11-397E7.4        SULT1A2             RP11-17E13.2       
[171] TCP11               HLA-B               BTN3A2              KLF14               XXbac-BPG300A18.13 
[176] AP3B2               LINC00310           TRIM73              CLUAP1              RP11-419C5.2       
[181] HSD17B13            KLHL31              RP1-313I6.12        OPRL1               SLC35G5            
[186] LRRC37A             SDHDP6              POM121C             RP11-378J18.8       SYPL2              
[191] HEY2                RNF5                KAT8                TMBIM1              TCEA3              
[196] HERPUD1             HLA-DRA             PRMT6               RP13-753N3.1        CCDC163P           
[201] AFF1                HMGN1               LACTB               PACS1               GUSBP4             
[206] RP11-64K12.4        LINC00202-1         CCDC36              SERPINC1            ARL14EP            
[211] NPIPA1              AMIGO1              LGALS9C             ROM1                MMP16              
[216] TIGD7               CNTN2               RPAP1               PDLIM4              PEMT               
[221] CWF19L1             INVS                CELSR2              MARVELD3            RP11-107F6.3       
[226] EYA1                TRMT61B             RP11-680G24.5       RP11-324E6.9        SUZ12P             
[231] PKD1                WARS2               LHCGR               AHSA2               LRRC37A2           
[236] RP5-874C20.3        AC138783.12         ANK1                SLC7A9              CCHCR1             
[241] TMEM60              ZNF589              LMOD1               MICA                RP11-502I4.3       
[246] NUDCD3              RNF157              ADAM1A              ADCY10P1            VPS53              
[251] TRIM66              MRPS18AP1           MYBPC3              MARK3               KIAA1161           
[256] GYPE                CTD-2260A17.3       ALMS1P              RPP25               RP11-680G24.4      
[261] TOM1L2              PLEC                NBPF1               TMEM9B-AS1          CPNE1              
[266] HLA-DQA2            NFKBIL1             RP3-465N24.5        IL20RB              UGT1A7             
[271] AL022393.7          ZSCAN31             UGT1A9              UGT1A8              UGT1A3             
[276] UGT1A5              RP11-114H24.5       HCAR2               ZC3H12C             IRS1               
[281] ELP6                RAN                 C18orf8             SLC9B1              LONRF1             
[286] FAM106A             OXCT2               SMG5                RP11-493E12.2       TTC12              
[291] PHC1                NEURL2              UGT1A1              TRIM24              RMDN1              
[296] CEACAM21            TLK1                UGT1A10             PDXDC2P             GS1-259H13.2       
[301] LRRC36              SFI1                SPATA5L1            DHRS11              EIF3CL             
[306] OXCT2P1             TSPY26P             MRPL45P2            RP5-966M1.6         LPIN3              
[311] RP11-593F23.1       RP11-339B21.14      LINC00674           UGT1A6              REEP2              
[316] UGT1A4              SIL1                LRRC45              ABLIM3              NDUFS1             
[321] GPIHBP1             APOC1P1             PLTP                RP11-1348G14.1      BEND6              
[326] SAA3P               SULT1A1             HLA-DRB5            SURF1               RAC1               
[331] ADIPOQ-AS1          EFCAB13             FIGNL1              HCAR1               STAG3L2            
[336] AC131056.3          KNOP1               CYP17A1             CTNNAL1             CTC-510F12.4       
[341] NPPA-AS1            GPR180              TEX264              RP11-304L19.3       MT1E               
[346] MST1R               RP11-817O13.8       MKX                 GOSR1               RPL8               
[351] ANXA5               GALNT2              RP1-265C24.5        PARP6               TCF19              
[356] SSC5D               ENDOG               RP11-196G11.2       PCNX                AF131215.9         
[361] GATSL3              PRPH2               FGF9                PTH1R               ZNF577             
[366] FAM182A             TTYH1               KHK                 RP11-457M11.5       RMI2               
[371] SPC25               AC007390.5          RP1-130H16.18       CAMK2G              PDE3A              
[376] TMEM116             TNXA                RP11-573D15.1       CTA-85E5.10         AC138430.4         
[381] DNAH17              RP11-408A13.4       C4B                 OLFM2               NUP85              
[386] PTPN12              CYP17A1-AS1         B3GALNT2            RP1-18D14.7         RSPO3              
[391] RP11-24N18.1        LRRC37A6P           GATM                NAP1L4              RP11-7F17.5        
[396] RP11-433P17.3       DAP3                CYBRD1              TMEM110             PABPC4             
[401] RP5-968P14.2        C1orf167            GFPT1               RP11-345M22.2       AC010883.5         
[406] AC012146.7          RP11-10L12.4        CROCCP2             RP11-111A22.1       LINC00886          
[411] AC006129.4          GRK4                TBKBP1              CMIP                ZNF781             
[416] MST1                CCL3L3              YPEL3               HAPLN4              SKIV2L             
[421] ZNF703              RFC4                PEG3                UFSP1 
\end{verbatim}
\end{footnotesize}     

\bibliographystyle{chicago}
\bibliography{eGene_ref}

\end{document}